# Negligible oxygen vacancies, low critical current density, electric-field modulation, in-plane anisotropic and high-field transport of a superconducting $Nd_{0.8}Sr_{0.2}NiO_2/SrTiO_3$ heterostructure

Xiaorong Zhou[#], Zexin Feng[#], Peixin Qin, Han Yan, Xiaoning Wang, Pan Nie, Haojiang Wu, Xin Zhang, Hongyu Chen, Ziang Meng, Zengwei Zhu, Zhiqi Liu*

**Abstract** The emerging Ni-based superconducting oxide thin films are rather intriguing to the entire condensed matter physics. Here we report some brief experimental results on transport measurements for a 14-nm-thick superconducting $Nd_{0.8}Sr_{0.2}NiO_2/SrTiO_3$ thin-film heterostructure with an onset transition temperature of ~9.5 K. Photoluminescence measurements reveal that there is negligible oxygen vacancy creation in the $SrTiO_3$ substrate during thin-film deposition and post chemical reduction for the $Nd_{0.8}Sr_{0.2}NiO_2/SrTiO_3$ heterostructure. It was found that the critical current density of the $Nd_{0.8}Sr_{0.2}NiO_2/SrTiO_3$ thin-film heterostructure is relatively small, ~$4\times10^3$ A·cm$^{-2}$. Although the surface steps of $SrTiO_3$ substrates lead to an anisotropy for in-plane resistivity, the superconducting transition temperatures are almost the same. The out-of-plane magnetotransport measurements yield an upper critical field of ~11.4 T and an estimated in-plane Ginzburg-Landau coherence length of ~5.4 nm. High-field magnetotransport measurements up to 50 T reveal anisotropic critical fields at 1.8 K for three different measurement geometries and a complicated Hall effect. An electric field applied via the $SrTiO_3$ substrate slightly varies the superconducting transition temperature. These experimental results could be useful for this rapidly developing field.

X.-R. Zhou, Z.-X. Feng, P.-X. Qin, H. Yan, X.-N. Wang, H.-J. Wu, X. Zhang, H.-Y. Chen, Z.-A. Meng, Z.-Q. Liu*
School of Materials Science and Engineering, Beihang University, Beijing 100191, China
Email: zhiqi@buaa.edu.cn

P. Nie, Z.-W. Zhu
Wuhan National High Magnetic Field Center, and School of Physics, Huazhong University of Science and Technology, Wuhan 430074, China

Xiao-Rong Zhou and Ze-Xin Feng contributed equally to this work.

## 1 Introduction

The recent observation of superconductivity in $Nd_{0.8}Sr_{0.2}NiO_2$ thin films has attracted explosive interests due to the possible similarity of electronic structures between the nickelates and high-temperature cuprate superconductors [1]. However, it is found that the parent compound $NdNiO_2$ of the superconducting $Nd_{0.8}Sr_{0.2}NiO_2$ lacks in any magnetic order down to 1.7 K [2], which is different from superconducting cuprates where the antiferromagnetic ordering of parent compounds is important to high temperature superconductivity [3]. Hence, the superconductivity in $Nd_{0.8}Sr_{0.2}NiO_2$ thin films could be a new type of superconducting phenomenon that will help us understand the nature of the high-temperature superconductivity better.

Although there are various theoretical works on this new superconducting system [4-30], only a few experimental results were reported. As far as we have noticed, there are only four groups claiming successful synthesis of nanoscale superconducting $Nd_{0.8}Sr_{0.2}NiO_2$ thin films. Specifically, most of the experimental work [1, 31-34] in this area came from Prof. Harold Hwang's group at Stanford University who first realized this phenomenon. Additionally, Gu *et al.* [35] and Xiang *et al.* [36] from Prof. Haihu Wen's group at Nanjing University studied the superconducting gaps via scanning tunneling microscopy spectroscope and physical properties by transport measurements. Gao *et al.* [37] from Dr. Xingjiang Zhou's group at Institute of Physics, Chinese Academy of Sciences, reported the successful growth of superconducting infinite-layer $Nd_{0.8}Sr_{0.2}NiO_2$ thin films by chemical reduction in a quartz tube. Moreover, Zeng *et al.* [38] from Prof. Ariando's group at National University of Singapore achieved the superconductivity in $Nd_{0.8}Sr_{0.2}NiO_2$ thin films by reducing $Nd_{0.8}Sr_{0.2}NiO_3$ thin films with $CaH_2$ powder in a vacuum chamber at 340-360 °C for 80-120 min. Especially, the chemical reduction carried out in a pulse laser deposition vacuum chamber by Zeng *et al.* [38] could be much more feasible as most oxide electronics groups lack in experimental experience in dealing with chemical duction in Pyrex glass or quartz tube.

## 2 Experimental

Based on the chemical reduction process reported by Zeng *et al.* [38] and our previous experimental results [39], we have successfully fabricated superconducting Nd$_{0.8}$Sr$_{0.2}$NiO$_2$ films. Specifically, epitaxial 113-phase perovskite Nd$_{0.8}$Sr$_{0.2}$NiO$_3$ thin films were grown on (001)-oriented TiO$_2$-terminated SrTiO$_3$ single-crystal substrates by a pulsed laser deposition system at 700 ºC and 20 Pa oxygen partial pressure. Subsequently, Nd$_{0.8}$Sr$_{0.2}$NiO$_3$ samples were contained in Al foils with each piece mixed with 0.1 g CaH$_2$ powder and then moved to the pulsed laser deposition vacuum chamber. The chemical reduction was performed at 350 ºC for 2 h. During this chemical reduction, the vacuum chamber was sealed without connecting to the turbo pump. The ramping rate for heating and cooling was kept at 10 ºC·min$^{-1}$.

In this article, we report photoluminescence measurement results and the transport properties of a 14-nm-thick Nd$_{0.8}$Sr$_{0.2}$NiO$_2$/SrTiO$_3$ heterostructure, which includes critical current density, electric-field modulation, in-plane anisotropy and high-field magnetotransport properties of such a superconducting system. We noticed that all these aspects have not been reported in previous experimental studies [1,31-38].

## 3 Results and discussion

Figure 1 shows the photoluminescence spectra of the Nd$_{0.8}$Sr$_{0.2}$NiO$_2$/SrTiO$_3$ heterostructure and a bare SrTiO$_3$ substrate that was purchased from Crystec and used to fabricate Nd$_{0.8}$Sr$_{0.2}$NiO$_2$/SrTiO$_3$ heterostructures. Oxygen vacancies in SrTiO$_3$ typically lead to multiple defect photoluminescence peaks at a wavelength range between 350 and 600 nm and the intensity of the peaks are very sensitive to the concentration of oxygen vacancies [40,41]. Accordingly, the photoluminescence spectra could be used to qualitatively examine the oxygen vacancy concentration on the surface region of a bulk SrTiO$_3$ single crystal. Compared with the photoemission signal of the bare SrTiO$_3$ substrate that is supposed to possess only a tiny amount of oxygen vacancies, the defect-related in-gap photoemission of the superconducting Nd$_{0.8}$Sr$_{0.2}$NiO$_2$/SrTiO$_3$ heterostructure becomes even weaker. This indicates that oxygen vacancies in the superconducting Nd$_{0.8}$Sr$_{0.2}$NiO$_2$/SrTiO$_3$ heterostructure are negligible.

As the 113-phase Nd$_{0.8}$Sr$_{0.2}$NiO$_3$/SrTiO$_3$ heterostructure was fabricated at a high oxygen partial pressure of 20 Pa at 700°C in our pulsed laser deposition system, the deposition process could act as oxygen annealing for the SrTiO$_3$ substrate used for the Nd$_{0.8}$Sr$_{0.2}$NiO$_3$ film fabrication, which can thus lower the photoluminescence peak. On the other hand, after chemical reduction, the conducting film capping could suppress the photoluminescence emission as well due to the absorption of photons by free electrons. In addition, the weaker photoemission intensity of the superconducting Nd$_{0.8}$Sr$_{0.2}$NiO$_2$/SrTiO$_3$ heterostructure relative to the bare SrTiO$_3$ substrate suggests that the post chemical reduction process for achieving the 112-phase Nd$_{0.8}$Sr$_{0.2}$NiO$_2$/SrTiO$_3$ heterostructure does not produce noticeable oxygen vacancies as well. This experimental aspect could be rather useful to theoretically model the superconducting Nd$_{0.8}$Sr$_{0.2}$NiO$_2$/SrTiO$_3$ heterostructure.

The temperature-dependent resistivity of the Nd$_{0.8}$Sr$_{0.2}$NiO$_2$/SrTiO$_3$ heterostructure measured by the linear four-probe geometry is plotted in Fig. 2a. It shows metallic behavior at high temperature with a room-temperature resistivity of ~1.8 mΩ·cm, which is comparable with that reported in Ref. [1]. The superconducting phase transition occurs at an onset transition temperature of ~9.5 K and a zero-resistance state arises at ~3.0 K. Figure 2b plots current-density-dependent resistance measured at different temperatures. The critical current density of the superconducting Nd$_{0.8}$Sr$_{0.2}$NiO$_2$ film is ~4×10$^3$ A·m$^{-2}$, which is much smaller than that of some typical superconductors such as Nb (~10$^6$ A·m$^{-2}$) [42], YBa$_2$Cu$_3$O$_{7-x}$ (~8×10$^6$ A·m$^{-2}$) [43], MgB$_2$ (~10$^6$ A·m$^{-2}$) [44] and Fe-based superconductors (~2×10$^6$ A·m$^{-2}$) [45]. The critical current density of a superconductor is typically associated with the magnetic flux pinning. The relatively small value of the critical current density, corresponding to a weak pinning capability, could be an intrinsic property of superconducting Nd$_{0.8}$Sr$_{0.2}$NiO$_2$ thin films. Alternatively, it may be related to the small thickness of the sample as a result of interfacial/surface scattering.

Up to now, the emerging Ni-based oxide superconductivity has not been achieved in bulk nickelates [46-48] such as Nd$_{0.8}$Sr$_{0.2}$NiO$_2$, which suggests that the surface or interfacial effects between the films and substrates may play an important role in this new superconducting phenomenon. It is worth emphasizing that our thin films were fabricated on buffered HF-treated SrTiO$_3$ substrates that exhibit single TiO$_2$ terminations. Owing to the steps on the surfaces of TiO$_2$-terminated SrTiO$_3$ substrates as a result of a vicinal miscut angle, we specifically investigated the effect of surface steps on the electrical transport properties of the Nd$_{0.8}$Sr$_{0.2}$NiO$_2$/SrTiO$_3$ thin-film heterostructure.

Our atomic force microscopy measurements reveal that the surfaces steps of TiO$_2$-terminated SrTiO$_3$ substrates are almost parallel to one in-plane edge of the (001) plane. Therefore, we made linear four-probe electrical contacts with an Al wire bonder along two orthorhombic in-plane edges. The measurement geometry is schematized in Fig. 3a. As shown in Fig. 3b, the temperature-dependent resistivity for different measurement geometries exhibits anisotropic behavior and the room-temperature resistivity for the current direction perpendicular to SrTiO$_3$ surface steps is ~7.7% higher than that for the current direction parallel to SrTiO$_3$ surface steps. This suggests that the electron scattering at the surface steps can remarkably increase the resistivity. However, the onset transition temperature is ~9.5 and ~10.0 K for the current perpendicular and parallel to the surface steps, respectively, which are almost the same. It implies that although the surface steps could yield an intrinsic anisotropy in electrical transport of the normal state, they have a negligible effect for the superconducting state of the $Nd_{0.8}Sr_{0.2}NiO_2/SrTiO_3$ thin-film heterostructure.

Subsequently, we performed out-of-plane magnetotransport measurements with static magnetic fields up to 9 T for the $Nd_{0.8}Sr_{0.2}NiO_2/SrTiO_3$ thin-film heterostructure. The measurement geometry is illustrated in Fig. 4a. As shown in Fig. 4b, the superconductivity is suppressed by the increasing magnetic field but not completely destroyed up to 9 T, which indicates that the out-of-plane upper critical field is larger than 9 T. Furthermore, the half normal-state resistivity value was utilized to estimate the upper critical field $\mu_0 H_{c2}(0\ K)$. As illustrated in Fig. 4c, the upper critical field at 0 K can be estimated based on the Werthamer-Helfand-Hohenberg formula [49] $\mu_0 H_{c2}(0\ K) = -0.7 T_c \left(\frac{d\mu_0 H_{c2}}{dT}\right)_{T_c}$. As a result, the zero-temperature upper critical field is obtained to be ~11.37 T. According to the Ginzburg-Landau equation [50] $\mu_0 H_{c2}(0\ K) = \frac{\Phi_0}{2\pi \xi_{GL}^2(0\ K)}$, where $\Phi_0$ is magnetic flux quantum and $\xi_{GL}$ is the Ginzburg-Landau coherence length, we obtain an in-plane Ginzburg-Landau coherence length of ~5.4 nm.

We notice that high-magnetic-field transport properties of superconducting $Nd_{0.8}Sr_{0.2}NiO_2/SrTiO_3$ thin-film heterostructures have not been reported either due to the lack of superconducting samples or limited high-field experiment equipment. To fill this research gap, we collaborated with an experimental group at Wuhan National High Magnetic Field Center, Huazhong University of Science and Technology [51] and performed systematic pulsed high-field magnetotransport measurements up to 50 T for the $Nd_{0.8}Sr_{0.2}NiO_2/SrTiO_3$ thin-film heterostructure.

As shown in Fig.5a, with the pulsed magnetic field increasing, the resistivity of the superconducting state at 1.8 K increases from zero to a finite value. At ~11.5 T, the normal-state resistivity is achieved and afterwards the resistivity trends to saturate. For the normal state at 20 K (Fig. 5b), a small negative magnetoresistance exists for low fields. However, the magnetoresistance trends to become positive for high-fields, likely due to the regular Lorentz-force-related orbital scattering. As the weak localization related negative magnetoresistance is typically observed in two-dimensional electron systems, the negative magnetoresistance could be possibly be related to magnetic spins in the normal state of the $Nd_{0.8}Sr_{0.2}NiO_2/SrTiO_3$ thin-film heterostructure.

For a thin-film superconductor, it is always interesting to examine whether the critical magnetic field to break the superconducting state is isotropic for out-of-plane and in-plane magnetic fields. Regarding this issue, it seems that different groups have acquired slightly different results. For example, the experimental research performed by the Stanford group [34] suggested that the upper critical field is isotropic for out-of-plane and in-plane magnetic fields, while the study conducted by the Nanjing University group [36] implied that the upper critical field is larger when the magnetic field is applied in plane.

To clarify this issue, we performed high-field magnetoresistance measurements for three different geometries at the zero-resistance state 1.8 K. As shown in Fig. 6, when the magnetic field is applied out of plane, the critical field for achieving the saturated normal-state resistivity is ~11.5 T, while it is ~14.5 T for the magnetic field applied in plane regardless of the in-plane angle between the magnetic field and the measuring current. The ~26% difference in the critical fields reflects that the out-of-plane magnetic field is easier to fully break the superconducting state via the Lorentz-force-related orbital scattering. In contrast, when the magnetic field is applied in plane, the orbital scattering is somehow suppressed for a nanoscale thin film sample, and thus the upper critical field based on the WHH theory that only considers the orbital effect is larger. However, both the upper critical fields along in-plane and out-of-plane are smaller than the Paulli limit ~17.5 T. Our experimental results agree well with what the Nanjing University group has concluded [36].

From the first report [1], the emerging Ni-based oxide superconductor exhibits multiple-band transport as its Fermi surface involves both electron and hole pockets. To check it out, we performed pulsed high-field Hall measurements up to 50 T as well. The specific electrical connections are illustrated in Fig. 7a. The collected Hall voltage signals versus the magnetic field are plotted in Fig. 7b for different temperatures for the normal state above 10 K. Above 20 K, the Hall effect pertains to the *n*-type electron transport considering our measurement geometry in Fig. 7a and gradually becomes nonlinear at fields higher than ~20 T. At 20 K, the Hall effect is complicated by multiple curvature changes. Below 20 K, the Hall effect below 15 T corresponds to the *p*-type hole transport, but it changes significantly at higher fields later. All these data depict that the Fermi surface of the $Nd_{0.8}Sr_{0.2}NiO_2/SrTiO_3$ thin-film heterostructure containing different types of pockets evolves with temperature and magnetic field.

The ground states of strongly correlated oxides are closely related to the carrier density. Accordingly, we further explored how the superconducting transition could possibly evolve with a perpendicular gating electric field applied onto the $SrTiO_3$ substrate, which can inject electrons/holes into the conducting $Nd_{0.8}Sr_{0.2}NiO_2$ thin-film channel. As $SrTiO_3$ has a rather large dielectric constant at low temperatures [52], the electrostatic tuning could be most effective for low temperatures, which is different from the piezoelectric strain modulation operated at room temperature [53-63]. As a result, we have examined the temperature-dependent resistivity below 12 K for different gating electric fields ($E_G$). As shown in Fig. 8, a negative electric field of -4 kV·cm$^{-1}$ that injects holes into the conducting thin-film channel slightly increases the superconducting transition temperature, while a positive electric field of +3.6 kV·cm$^{-1}$ decreases the onset superconducting temperature by ~0.5 K. Such a modulation is quite similar to the electric-field effect in superconducting YBa$_2$Cu$_3$O$_7$ films [64]. Therefore, it is feasible to tune the superconducting phase transition of the Ni-based oxide superconductivity by electrostatic modulation, which could be further enhanced via ionic liquid modulation.

## 4 Conclusion

Up to now, the superconductivity has not been realized in bulk infinite-layer nickelates. If later possibly achieved, the critical current densities of bulk Ni-based superconductors could be much higher than the values we observed for thin-film heterostructure samples. Oxygen vacancies are negligible in the SrTiO$_3$ substrate of the superconducting Nd$_{0.8}$Sr$_{0.2}$NiO$_2$/SrTiO$_3$ heterostructure. The surface topography obviously affects the normal-state transport properties of the Nd$_{0.8}$Sr$_{0.2}$NiO$_2$/SrTiO$_3$ thin-film heterostructure but does not vary the superconducting parameters remarkably, which suggests that the superconducting properties of the nickelates could be, to some extent, independent of the normal-state electron scattering. The out-of-plane negative magnetoresistance effect for the normal resistivity state is observed, which could be related to magnetic spins in the Nd$_{0.8}$Sr$_{0.2}$NiO$_2$ thin film. In addition, the anisotropic critical fields for out-of-plane and in-plane magnetic fields could be a common feature for superconducting nanoscale thin-film heterostructures, simply due to the shape anisotropy and the resulting suppression of the orbital scattering. Our preliminary experiments indicate that the superconducting transition temperatures can be modulated by electrostatic doping. All these properties could be useful to design superconducting circuits [65]. In addition, it would be rather interesting to fabricate superconducting Nd$_{0.8}$Sr$_{0.2}$NiO$_2$ thin films onto ferromagnetic substrates [66] such as oxygen-deficient Nb-SrTiO$_3$ to investigate the interplay between magnetism and superconductivity. Finally, we hope that these experimental results could be helpful for understanding this intriguing electron system.

**Acknowledgments** This study was financially supported by the National Natural Science Foundation of China (Nos. 51822101, 51861135104 and 51771009).

**Figure 1**

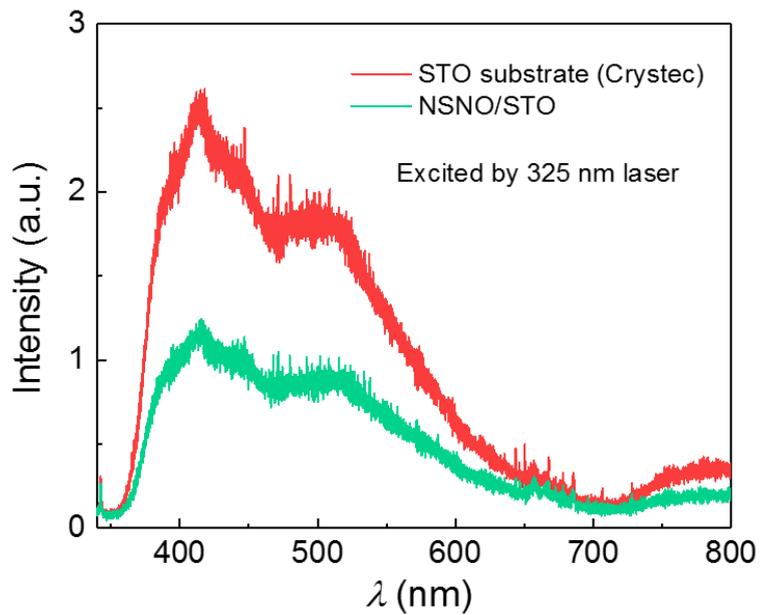

**Fig. 1** Room-temperature photoluminescence of a bare SrTiO$_3$ (STO) substrate purchased from Crystec and a superconducting Nd$_{0.8}$Sr$_{0.2}$NiO$_2$/SrTiO$_3$ (NSNO/STO) heterostructure excited by 325 nm laser.

**Figure 2**

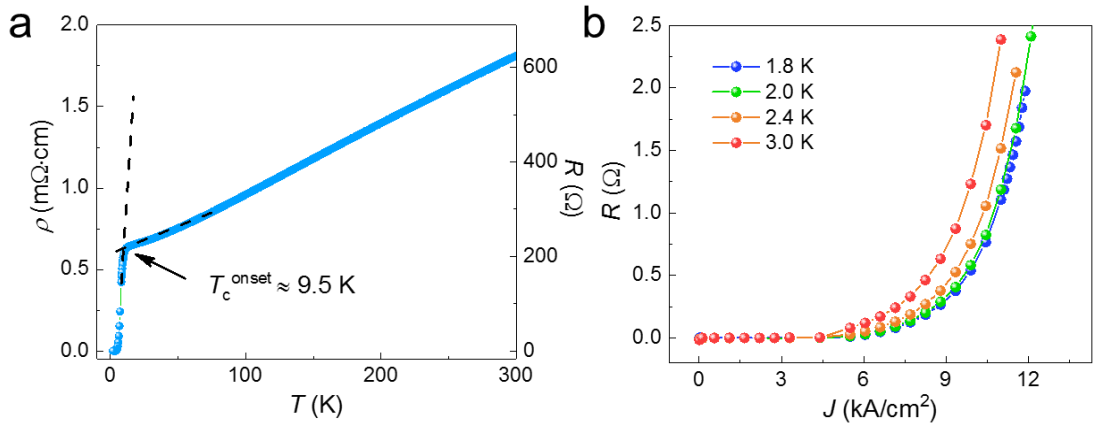

**Fig. 2 a** Temperature-dependent resistivity of a 14-nm-thick $Nd_{0.8}Sr_{0.2}NiO_2/SrTiO_3$ thin-film heterostructure; **b** current density dependent resistance of the $Nd_{0.8}Sr_{0.2}NiO_2/SrTiO_3$ heterostructure at different temperatures.

**Figure 3**

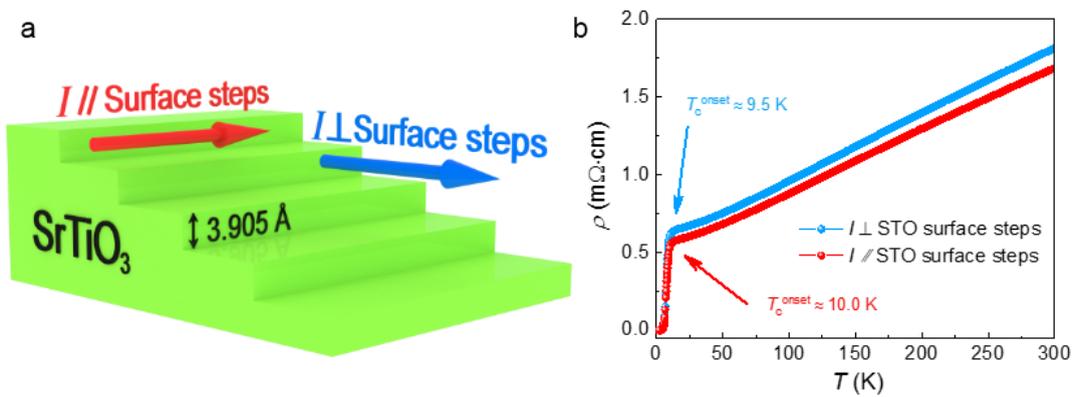

**Fig. 3 a** Schematic of the measurement geometry for different current directions; **b** temperature-dependent resistivity of the $Nd_{0.8}Sr_{0.2}NiO_2/SrTiO_3$ heterostructure measured by different current directions.

**Figure 4**

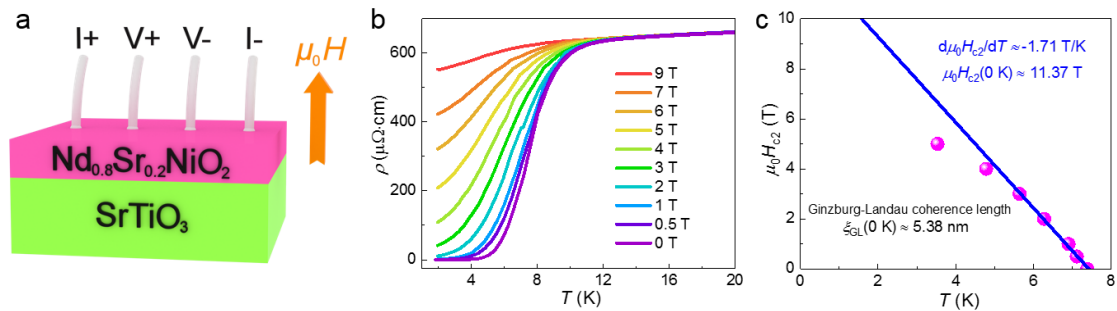

**Fig. 4 a** Schematic of the measurement geometry for the out-of-plane magnetotransport; **b** temperature-dependent resistivity of the $Nd_{0.8}Sr_{0.2}NiO_2/SrTiO_3$ heterostructure under different magnetic fields ranging from 0 to 9 T; **c** the upper critical field as a function of the temperature

**Figure 5**

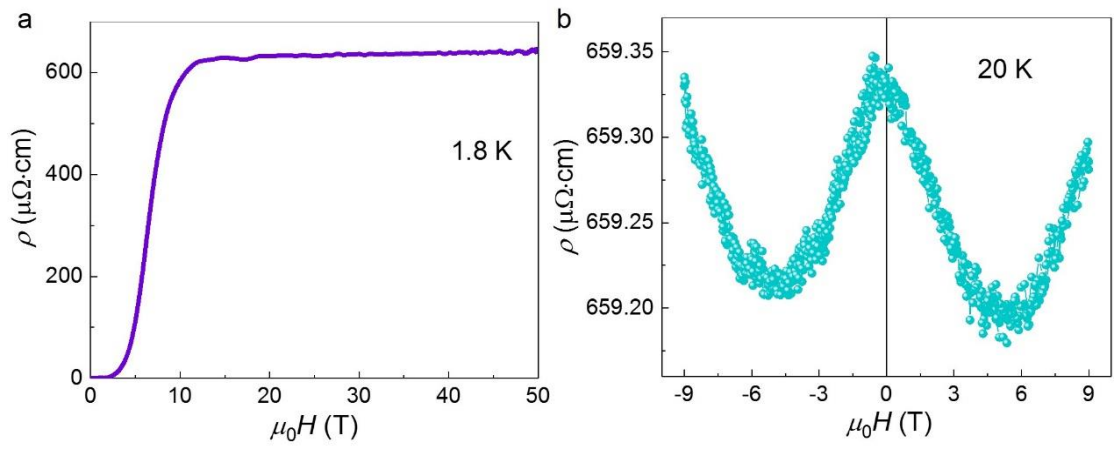

**Fig.5 a** Resistivity as a function of magnetic field up to 50 T at 1.8 K; **b** static-field magnetoresistance up to 9 T for the normal state at 20 K.

**Figure 6**

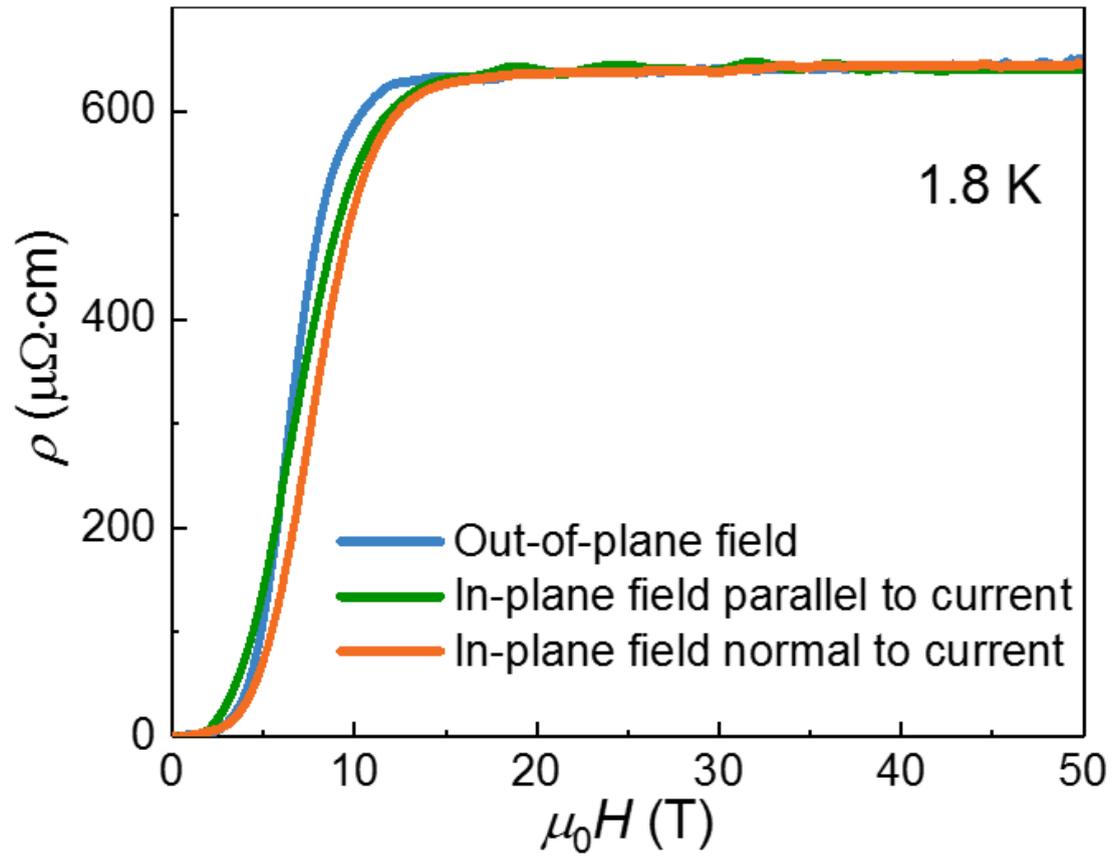

**Fig. 6** Resistivity as a function of magnetic field up to 50 T at 1.8 K for three measurement geometries between the measuring current and the magnetic field.

**Figure 7**

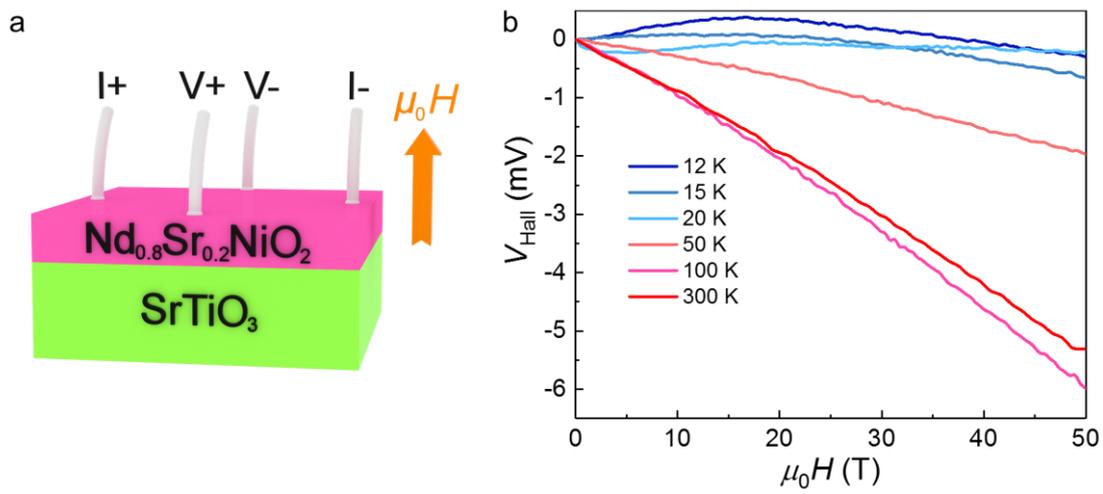

**Fig. 7 a** Cartoon illustration for the Hall measurement connections; **b** Hall voltage versus magnetic field up to 50 T for the normal state above 10 K.

Figure 8

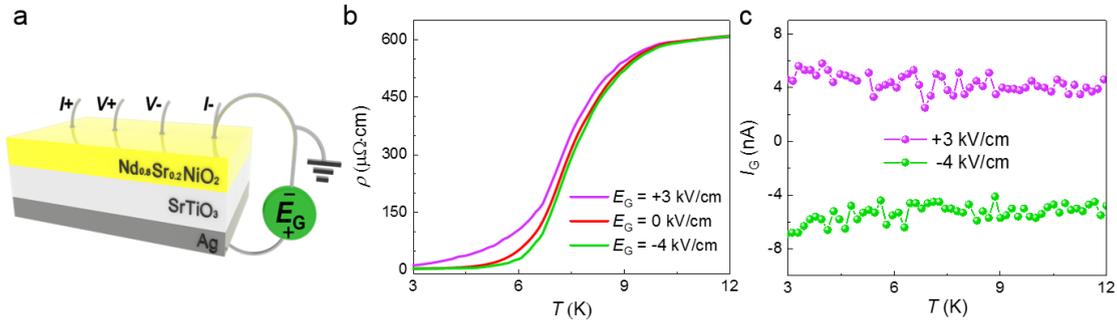

**Fig. 8 a** Schematic of the measurement geometry of the electric-field modulation experiment; **b** Superconducting transition for different gating electric fields $E_G$; **c** Perpendicular gating current versus temperature under different gating electric fields.